\documentclass[aps,preprint,showpacs,preprintnumbers,amsmath,amssymb]{revtex4}

\usepackage{graphicx}
\usepackage{dcolumn}
\usepackage{bm}
\usepackage{hyperref}

\newcommand{\beq}{\begin{equation}}
\newcommand{\eeq}{\end{equation}}
\newcommand{\beqa}{\begin{eqnarray}}
\newcommand{\eeqa}{\end{eqnarray}}
\newcommand{\beqar}{\begin{eqnarray*}}
\newcommand{\eeqar}{\end{eqnarray*}}
\newcommand{\bra}[1]{\mbox{$\left\langle{#1}\right|$}}
\newcommand{\ket}[1]{\mbox{$\left|{#1}\right\rangle$}}

\def\I{{\rm i}}

\newcounter{saveeqn}

\begin{document}
\title{The energy spectrum symmetry \\
of Heisenberg model in Fock space}
\author{An Min Wang and Rengui Zhu}
\affiliation{Department of Modern Physics and Institute for
Theoretical Physics\\ University of Science and Technology of
China, Hefei, 230026, P.R.China}

\begin{abstract}
We prove strictly that one dimension spin $1/2$ Heisenberg model has
a symmetry of energy spectrum between its subspace $n$ and the
subspace $L-n$ of the Fock space. Our proof is completed by
introducing two general quantum operations. One is a flip operation
of spin direction and another is a mirror reflection of spin sites.
\end{abstract}


\maketitle

Heisenberg model has being studied almost continuously since it was
introduced by Heisenberg in 1926. Heisenberg Hamiltonian is one of
the elementary, important and interesting quantum many-body
operators. Understanding its energy spectrum is a guiding problem in
both mathematics and physics, in special, the physics of strongly
correlated many-body quantum systems.

In this brief report, we focus our attention on the energy spectrum
symmetry of Heisenberg model in Fock space. This problem was
motivated by one of authors' previous work \cite{anmwang}, in which,
a direct diagonlization of the BCS pairing Hamiltionian \cite{bcs}
by the decomposition of spin Fock space was proposed, and such a
symmetry of energy spectrum between the subspace $n$ and subspace
$L-n$ of Fock space was found by the numerical calculation and
analytical continuation. However, a physical symmetry had better to
be a strict property and then we shall give a strict proof
analytically to it, moreover, our conclusion has been extended to a
more general case -- Heisenberg model here. This symmetry of energy
spectrum will decrease significantly the amount of computation and
understand helpfully some physical features of quantum many-body
systems including Heisenberg model.

Let's consider a general Hamiltonian of a spin $1/2$ system in one
dimension:
\begin{equation}\label{eq:a}
 H=H_0+H_1
\end{equation}with
\begin{equation}
H_0=\frac{1}{2}\sum_{m=1}^L\epsilon_m\sigma_z^{(m)}
\end{equation}and
\begin{equation}
H_1=-\frac{1}{2}\sum_i\sum_{m<l=1}^{L}V_{ml}^i\sigma_i^{(m)}\sigma_i^{(l)}
\end{equation}
In fact, it is a common form of Heisenberg model\cite{mahan}, where
$\sigma_i$ is the usual Pauli matrix with $i=x,y,z$, and the
coupling coefficient $V_{ml}^i=V_{lm}^i$ is real and symmetric. A
special example of this kind of Hamiltonian is the spin analogy of
the BCS pairing Hamiltonian \cite{lawu,our1} as the following
\begin{equation}
H_p^{(L)}=\frac{1}{2}\sum_{m=1}^L\epsilon_m\sigma_z^{(m)}-\frac{1}{2}\sum_{m<l=1}^LV_{ml}
\left(\sigma_x^{(m)}\sigma_x^{(l)}+\sigma_y^{(m)}\sigma_y^{(l)}\right)
\end{equation}
Here $L$ is the number of the pairs or qubits. Actually, in order to
study the quntum simulating of spin system \cite{our2}, it is
natural, convenient and useful to study the solutions of these
systems in their spin Fock space \cite{anmwang}.

Obviously, the Hamiltonian of Heisenberg model including the spin
analogy of pairing model is an operator in the Hilbert space
$S_{spin}^{(L)}$ which is made of the direct product of $L$ spin
$1/2$ spaces. In the representation of $\sigma_z$ diagonalization,
the basis vectors of each spin $1/2$ space can be denoted by the
spin-up state with $0$, and the spin-down state with $1$, that is,
\begin{equation}
|0\rangle=\begin{pmatrix}1\\
0\end{pmatrix},\quad |1\rangle=\begin{pmatrix}0\\
1\end{pmatrix}
\end{equation}
So the basis vectors in the total Hilbert space can be written as
$\ket{\alpha_1\alpha_2\cdots\alpha_L}$, where $\alpha_i$ takes $0$
or 1 and $\alpha_1\alpha_2\cdots\alpha_L$ can be thought of a
qubit-string distributed with $0$ and $1$. If we introduce the
spin-ladder operator
$\sigma^{(\pm)}=\displaystyle\frac{1}{2}(\sigma_x \pm i\sigma_y)$,
it's easy to see that \beqa
\sigma^{(+)}|0\rangle=0,\quad \sigma^{(+)}|1\rangle=|0\rangle\\
~\sigma^{(-)}|0\rangle=|1\rangle,\quad \sigma^{(-)}|1\rangle=0
\eeqa This implies that $|0\rangle$ can be regarded as a state
with a single fermion occupation, $|1\rangle$ is an empty state of
fermion, and then $\sigma^{(+)}\sigma^{(-)}$ is equivalent to a
fermion number operator.

In fact, the above analysis indicates that the spin space $S_{\rm
spin}^{(L)}$ in a $L$-pair (qubit) system can be divided into the
different subspaces of Fock space which correspond to the different
numbers of spin-up states, that is $S_{\rm
spin}^{(L)}=S_{0}^{(L)}\oplus S_{1}^{(L)}\oplus
S_{2}^{(L)}\oplus\cdots\oplus S_{L}^{(L)}$, where the subspace $n$,
$i.e$ $S_n^{(L)}$, is a subspace with $n$ spin-up states $\ket{0}$
(corresponding to ``occupation"), and its basis is denoted by
$\ket{s_{i_1i_2\cdots i_n}^{(L)}}$ where $i_1,i_2,\cdots,i_n$
indicate the values of the positions appearing $0$ in the bit-string
$\alpha_1\alpha_2\cdots\alpha_L\;(\alpha_i=0$ or $1$). In other
words, the vectors belonging to $S_n^{(L)}$ with the fermion number
is $n$. Obviously the dimension of $S_n^{(L)}$ is
$\displaystyle\frac{L!}{n!(L-n)!}$.

From
$\left(\sigma_x^{(m)}\sigma_x^{(l)}+\sigma_y^{(m)}\sigma_y^{(l)}\right)
=\left(\sigma_x^{(m)}+\I\sigma_y^{(m)}\right)
\left(\sigma_x^{(l)}-\I\sigma_y^{(l)}\right)=\sigma_{m}^{(+)}\sigma_{l}^{(-)},
(m\neq l)$ and $\sigma_z^{(m)}$ as well as
$\sigma_z^{(m)}\sigma_z^{(l)}$ is diagonal in the above Hamiltonian
(\ref{eq:a}), it follows that for the arbitrary basis
$\ket{s_{i_1\cdots i_n}^{(L)}}$ belonging to $S_n^{(L)}$,
$H\ket{s_{i_1\cdots i_n}^{(L)}}$ also belongs to $S_n^{(L)}$ because
that $\sigma^+$ and $\sigma^-$ appear in pairs or do not appear in
the various terms of $H$. It implies that $\bra{s_{i_1\cdots
i_{m}}^{(L)}}H\ket{s_{i_1^\prime\cdots i_n^\prime}^{(L)}}=0$, (If
$m\neq n; m,n=1,2,\cdots, L$) and $\bra{2^{L}}H_p\ket{s_{i_1\cdots
i_n}^{(L)}}=0$. Therefore, we have proved that the pairing model
Hamiltonian is able to be decomposed into the direct sum of
submatrices in Fock subspaces \cite{anmwang}, that is
\begin{equation}
H^{(L)}=H_{{\rm sub}0}^{(L)}\oplus H_{\rm sub1}^{(L)}\oplus H_{\rm
sub2}^{(L)}\oplus\cdots\oplus
H_{{\rm sub}L}^{(L)} \label{dirsum1}%
\end{equation}

In order to prove the symmetry property of energy spectrum of
Heisenberg model between the subspaces of the spin Fock space, we
need to introduce two useful quantum operations on the Fock space.
The first operation denoted with ${\mathcal{F}}$, is a flip
operation of spin direction, which changes 0 (up) to 1 (down) and 1
(down) to 0 (up) in the qubit-string:
\begin{equation}
{\mathcal{F}}\equiv\bigotimes_{i=1\rightarrow}^L\sigma_x^{(i)}
\end{equation}
Here, $\rightarrow$ means that the factors are arranged from left to
right corresponding to $i$ from $1$ to $L$. Obviously, we have
${\mathcal{F}}={\mathcal{F}}^{-1}$.

The second operation we introduce, denoted by ${\mathcal{M}}$, is a
mirror reflection of spin sites. Actually, it is a kind of
rearrangement of sites of qubit-string, that is, it reflects
$\alpha_m$ originally sited in the $m$-th position of a qubit-string
$\alpha_1\alpha_2\cdots\alpha_N$ to a new place sited in the
$(L+1-m)$-th position. This operator's effect can be illustrated as
placing a mirror vertical to the qubit-string. In order to give an
explicit form of ${\mathcal{M}}$, we first define the other two
operations of which ${\mathcal{M}}$ consists. One is a swapping
operation of two neighboring bit positions:
\begin{equation}
S(i,i+1)\equiv\frac{1}{2}\sum_{\mu=0}^3\sigma_{\mu}^{(i)}\sigma_{\mu}^{(i+1)}
\end{equation}
which exchanges the spin state (or the fermion occupation number)
of site $i$ and site $i+1$ in the qubit-string. Here $\sigma_0$
denotes the identity matrix. Another is a kind of rearranging
operation:
\begin{equation}
P(j, k)\equiv\prod_{i=j\leftarrow}^{k-1}S(i,i+1)
\end{equation}
which extracts out the spin-state of site $j$, and rearranges it to
the site $k$ in the qubit-string. Note that $\leftarrow$ means that
the factors are arranged from right to left corresponding to $i$
from $1$ to $L$. Now, in terms of $P(j,k)$, it's easy to write
${\mathcal{M}}$ as
\begin{equation}
{\mathcal{M}}\equiv\prod_{i=1\leftarrow}^{L-1}P(1,L+1-i)
\end{equation}
Obviously, we have ${\mathcal{M}}={\mathcal{M}}^{-1}$.

Now we can prove following fundamental relations:
\begin{equation}
{\mathcal{F}}\bigotimes_{i=1\rightarrow}^{L}|\alpha_i\rangle=\bigotimes_{i=1\rightarrow}^{L}|\alpha_i+1~{\rm
mod}~2\rangle
\end{equation}
\begin{equation}
{\mathcal{M}}\bigotimes_{i=1\rightarrow}^L|\alpha_i\rangle=\bigotimes_{i=1\leftarrow}^L|\alpha_i\rangle
\end{equation}
\begin{equation}
{\mathcal{M}}\sigma_i^{(m)}{\mathcal{F}}^{-1}=(-1)^{1+\delta_{ix}}\sigma_i^{(m)},
\quad {\mathcal{F}}\sigma^{(m)}{}^{(\pm)}{\mathcal{F}}^{-1}=\sigma^{(m)}{}^{(\mp)}\\
\end{equation}
\begin{equation}
{\mathcal{M}}\sigma_i^{(m)}{\mathcal{M}}^{-1}=\sigma_i^{(L+1-m)}
\end{equation}
It must be emphasized that ${\mathcal{F}}$'s action maps a state
belonging to the subspace $n$ to a state belonging to the subspace
$L-n$ in Fock space. While ${\mathcal{M}}$' action keeps a state
belonging to a given subspace still within its subspace.

Based on the relations above, it is easy to get the transformation
of Hamiltonian (\ref{eq:a}) under the two operations ${\mathcal{F}}$
and ${\mathcal{M}}$ respectively:
\begin{equation}\label{eq:d}
{\mathcal{F}}H_0{\mathcal{F}}^{-1}=-H_0
\end{equation}
\begin{equation}\label{eq:e}
{\mathcal{F}}H_1{\mathcal{F}}^{-1}=H_1
\end{equation}
\begin{equation}\label{eq:c}
{\mathcal{M}}H_0{\mathcal{M}}^{-1}=\frac{1}{2}\sum_{m=1}^L\epsilon_{L+1-m}\sigma_z^{(m)}
\end{equation}
\begin{equation}\label{eq:b}
{\mathcal{M}}H_1{\mathcal{M}}^{-1}=
-\frac{1}{2}\sum_i\sum_{m<l}^LV_{L+1-m,L+1-l}^i\sigma_i^{(m)}\sigma_i^{(l)}
\end{equation}
If we assume that $V_{ml}^i$ only depends on $|m-l|$ (constant is
regarded as an extreme case), then the equation (\ref{eq:b}) is
reduced to
\begin{equation}\label{eq:f}
{\mathcal{M}}H_1{\mathcal{M}}^{-1}=H_1
\end{equation}
Furthermore if we assume that $\epsilon_m$ is homogenously
distributed, \emph{i.e}
\begin{equation}
\epsilon_m=\epsilon_0+md
\end{equation}
then, we can get
\begin{equation}
\epsilon_{(L+1-m)}=\epsilon_0+(L+1-m)d=2\epsilon_0+(L+1)d-\epsilon_m
\end{equation}
So the equation(\ref{eq:c}) is rewritten as
\begin{equation}\label{eq:g}
{\mathcal{M}}H_0{\mathcal{M}}^{-1}=-H_0+\left[\epsilon_0+\frac{1}{2}(L+1)d\right]\sum_{m=1}^{L}\sigma_z^{m}
\end{equation}
From (\ref{eq:d}), (\ref{eq:e}), (\ref{eq:f}) and (\ref{eq:g}),
under the jointed operation of ${\mathcal{F}}$ and
${\mathcal{M}}$(one can prove that it's no difference who operates
first), we get:
\begin{equation}
H^\prime=({\mathcal{MF}})H({\mathcal{MF}})^{-1}
=H-\left[\epsilon_0+\frac{1}{2}(L+1)d\right]\sum_{m=1}^L\sigma_z^{(m)}
\end{equation}
This implies that the eigenvalue problem of the transformed
Hamiltonian can be related with one of the original Hamiltonian
under some preconditions.

Note the fact that $H^\prime -H$ is diagonal and its eigenvectors
belong to a certain subspace $S_n$ in the Fock space and the
corresponding degeneration is the dimensions of $S_n$. That is, for
an arbitrary state $\ket{\phi_n} \in S_n$, we have \beq
\sum_{m=1}^L\sigma_z^{(m)}\ket{\phi_n} = [n-(L-n)]\ket{\phi_n} \eeq
Consequently, if $H$'s eigenvectors also belong to $S_n$, we can
obtain the relation of eigenvalue problems between the transformed
Hamiltonian and the original Hamiltonian.

Actually, in the many interesting models we can see that
$V_{ml}^x=V_{ml}^y$ and they only depend on the distant $|m-l|$
(constant is regarded as an extreme case), for example, the
pairing model \cite{anmwang,bcs,mahan,lawu}. Thus, since
$\sigma_z=\sigma^{(+)}\sigma^{(-)}-\sigma^{(-)}\sigma^{(+)}
=2\sigma^{(+)}\sigma^{(-)}-1$,
 and $\sigma_x^{(m)}\sigma_x^{(l)}+\sigma_y^{(m)}\sigma_y^{(l)}
 =2(\sigma_m^{(+)}\sigma_l^{(-)}+\sigma_m^{(-)}\sigma_l^{(+)})$,
we can get the eigenvector of $H$ fully belonging to a certain
subspace $S_n$. Furthermore, denoting the $k$-th eigenvector in this
given subspace with $|\psi_n^k\rangle$, we have
$\sum_{m=1}^{L}\sigma_z^{(m)}{\mathcal{MF}}|\psi_n^k\rangle=
[(L-n)-n]{\mathcal{MF}}|\psi_n^k\rangle$.

Therefore we have proved the following theorem:

{\bf Theorem}: If the Hamiltonian of Heisenberg model has the form:
\begin{equation}
H=\frac{1}{2}\sum_{m=1}^L\sigma_z^{(m)}-\frac{1}{2}\sum_{m<l=1}^LV_{ml}
(\sigma_x^{(m)}\sigma_x^{(l)}+\sigma_y^{(m)}\sigma_y^{(l)})-\frac{1}{2}
\sum_{m<l=1}^LU_{ml}\sigma_z^{(m)}\sigma_z^{(l)}
\end{equation}
on the condition that $\epsilon_m=\epsilon_0+md$ and $V_{ml}$,
$U_{ml}$ depend only on $|m-l|$ (constant is regarded as an extreme
case), then, if $|\psi_n^k\rangle$ is one of its eigenvectors in
subspace $n$, ${\mathcal{MF}}|\psi_n^k\rangle$ must be its another
eigenvector in subspace $L-n$, moreover, with the eigenvalue
increased by
$[2\epsilon_0+d(L+1)]\left(\displaystyle\frac{L}{2}-n\right)$; and
this difference is independent of $k$.

If we set $\epsilon_0=0$, then it is just the formula of symmetry
of energy spectrum found in ref\cite{anmwang} by numerical
analysis, but it has been extended to a more general case with an
additive $\sigma_z$-$\sigma_z$ interaction.

In the end, we would like to point out that this strictly analytical
proof about the symmetry of energy spectrum of Heisenberg model in
the subspace $n$ and the subspace $L-n$ of Fock space will
significantly decrease the calculations about the spectrum of
Heisenberg model including the spin analogy of pairing model,
moreover this proof also accounts for that the direct
diagonalization of Fock space is useful and our energy spectrum data
obtained by this method is exact enough in the ref. \cite{anmwang}.
Further, we hope that our result can be helpful for understanding
quantum many-body systems.

We are grateful to Feng Xu, Xiaoqiang Su for helpful discussions,
and to Hao You, Ningbo Zhao, Xiaodong Yang, Wanqing Niu, Xiaosan Ma,
Dehui Zhan, Liang Qiu, Xuechao Li and Zhuqiang Zhang for our
cooperations in the group of quantum theory in the institute for
theoretical physics of University of Science and Technology of
China.  This work was founded by the National Fundamental Research
Program of China with No. 2001CB309310, partially supported by the
National Natural Science Foundation of China under Grant No.
60573008.

\end{document}